\documentclass[11pt,a4paper]{article}
\evensidemargin=\oddsidemargin
\def\squarebox#1{\hbox to #1{\hfill\vbox to #1{\vfill}}}
\newcommand{\qed}{\hspace*{\fill}
            \vbox{\hrule\hbox{\vrule\squarebox{.667em}\vrule}\hrule}\smallskip\newline}
\newtheorem{THEOREM}{Theorem}
\newenvironment{theorem}{\begin{THEOREM} \hspace{-.85em} {\bf :} \rm}                        {\end{THEOREM}}
\newtheorem{LEMMA}[THEOREM]{Lemma}
\newenvironment{lemma}{\begin{LEMMA} \hspace{-.85em} {\bf :} \rm}                      {\end{LEMMA}}
\newtheorem{COROLLARY}[THEOREM]{Corollary}
\newenvironment{corollary}{\begin{COROLLARY} \hspace{-.65em} {\bf :} \rm}                          {\end{COROLLARY}}
\newtheorem{FACT}[THEOREM]{Fact}

\newenvironment{proof}{\noindent {\bf Proof:} \hspace{-.2em}} {}

\newtheorem{DEFINITION}{Definition}
\newenvironment{definition}{\begin{DEFINITION} \hspace{-.85em} {\bf :} \rm}
                            {\end{DEFINITION}}
\newtheorem{PROPOSITION}{Proposition}

\newtheorem{CLAIM}[THEOREM]{Claim}
\newenvironment{claim}{\begin{CLAIM} \hspace{-.85em} {\bf :} \rm}
                      {\end{CLAIM}}
\usepackage{graphicx}

\usepackage{subfigure}
\usepackage{graphicx}
\usepackage{graphics}
\usepackage{amssymb}
\input{epsf}

\usepackage{algorithm}
\usepackage{algpseudocode}
\begin{document}
\title{Algorithms for the minimum non-separating path and the balanced connected bipartition problems on grid graphs\\
(With errata)}
\author{Bang Ye Wu\thanks{National Chung Cheng University, ChiaYi, Taiwan 621,
R.O.C., E-mail: bangye@cs.ccu.edu.tw}
\\
{\small Dept. of Computer Science and Information Engineering}\\
{\small National Chung Cheng University, Taiwan}}
\date{}
\maketitle
\subsection*{\centering Abstract}
{\em
For given a pair of nodes in a graph, the minimum non-separating path problem looks for a minimum weight path between the two nodes such that the remaining graph after removing the path is still connected.
The balanced connected bipartition (BCP$_2$) problem looks for a way to bipartition a graph into two connected subgraphs with their weights as equal as possible.
In this paper we present an algorithm in time $O(N\log N)$ for finding a minimum weight non-separating path between two given nodes in a grid graph of $N$ nodes with positive weight.
This result leads to a 5/4-approximation algorithm for the BCP$_2$ problem on grid graphs, which is the currently best ratio achieved in polynomial time.
We also developed an exact algorithm for the BCP$_2$ problem on grid graphs. 
Based on the exact algorithm and a rounding technique, we show an approximation scheme, which is a fully polynomial time approximation scheme for fixed number of rows.   }
{\flushleft\bf Key words.} algorithm, approximation algorithm, non-separating path, balanced connected partition, grid graphs.

\section*{About this version}
This article was published as \cite{wu13}. 
In this version we report a mistake about the \emph{minimum non-separating path} on \emph{grid graphs} and fix the 5/4-approximation algorithm for \emph{Balanced Connected 2-Partition} problem on grid graphs (GBCP$_2$).
The correction is at the appendix.

\section{Introduction}

Let $G=(V,E,w)$ be a connected undirected graph, in which $w$ is a nonnegative node weight function. 
For two given nodes $s$ and $t$, if we want to allocate some of the nodes for the communication between $s$ and $t$, choosing a minimum $st$-path (a minimum weight path with endpoints $s$ and $t$) may be the best way. However, if the chosen nodes cannot be used for other services, the remaining network may be separated into species.
To keep the remaining network connected, one may hope to find a minimum \emph{non-separating} $st$-path, i.e.,  
a $st$-path $P$ such that the remaining graph $G-P$ is connected.
However, in general, non-separating path does not always exist. 
A natural relaxation allows any connected subgraph containing both $s$ and $t$. That is, we look for a minimum weight connected subgraph $B$ containing both $s$ and $t$ such that $G-B$ is connected. We name such a subgraph by ``non-separating $st$-connector'' ($st$-NSC or simply ``NSC'' if no confusion), and the one with minimum weight is a  \emph{minimum non-separating connector} (min-NSC). 

A node bipartition $(U,V-U)$ is a \emph{connected bipartition} if both the subgraphs induced by $U$ and $V-U$ are connected. Immediately if $B$ is an NSC, then $(V(B),V-V(B))$ is a connected bipartition.
The maximum balance connected bipartition (BCP$_2$) problem looks for a connected bipartition $(U,V-U)$ such that the balance, defined by $\min \{w(U),w(V-U)\}$, is maximized, in which $w(U)$ denotes the total weight of nodes in $U$.
The applications of BCP may appear in image processing, data bases, operating
systems, cluster analysis, etc. \cite{cha07}.
An $m\times n$ grid graph $M$ is an undirected graph and can be thought of as a 2-dimensional matrix, in which $m$ and $n$ are the numbers of rows and columns, respectively.
The node set of $M$ can be represented by $V=\{M_{ij}| 1\leq i\leq m, 1\leq j\leq n\}$ and there exists an edge between two consecutive nodes in the same row or the same column.  
In this paper we study the min-NSC and the BCP$_2$ problems on node-weighted grid graphs.

The non-separating path problem has been studied from the perspective of graph theory. 
Most of the works are devoted to its relationship to graph connectivity \cite{bol96,che03,kaw05,kaw08,tut63}, but we haven't found any optimization problem about it. 
The min-NSC problem on general graphs is NP-hard in the strong sense and cannot be approximated with ratio
$|V|^{1-\varepsilon}$ for any $\varepsilon >0$ in polynomial time unless P=NP \cite{wu09a} (named ``minimum border problem''). 
In this paper we show that 
a minimum $st$-NSC on a grid graph is a minimum non-separating $st$-path and can be found in $O(N\log N)$ time, in which $N$ is the number of nodes.
The efficient algorithm is based on two key points. First, the min-NSC on a grid graph is a minimum weight path with at most one boundary subpath; and secondly, such a path can be found by reducing to a {\em range minimum query} (RMQ) problem. 

The second result of this paper is about the BCP$_2$ problem on grid graphs (GBCP$_2$ for short). 
Based on NSC and $st$-numbering, we propose a 5/4-approximation algorithm with time complexity $O(N\log N)$ for the GBCP$_2$ problem, which is the currently best result achieved in polynomial time. 
We also developed an exact algorithm for GBCP$_2$. For an $m\times n$ grid graph of total weight $W$, $m\leq n$, the algorithm takes $O(mNW8^m)$ time, which is more efficient than the naive brute force method of $O(N2^N)$ time. 
The exact algorithm uses a typical dynamic programming strategy and computes the best bipartition for any possible weight and any connection topology of the first $i$ columns for $i$ from 1 to $n$. The analysis itself is of its own interest. An obvious upper bound of the number of connection topologies is $m^m$. With a more precise analysis and using the method of generating function, we show a sharper bound of $O(2^m)$.
Based on the exact algorithm and a rounding technique, we developed an approximation scheme. For any $\varepsilon>0$, the GBCP$_2$ problem can be $(1+\varepsilon)$-approximated in $O((1+\frac{1}{\varepsilon})mN^28^m)$ time, which is a {\em fully polynomial-time approximation scheme} (FPTAS) for fixed $m$. 

The BCP$_q$ problem is a generalization of BCP$_2$, for which the input graph is partitioned into $q$ connected subgraphs for any given $q\geq 2$.
Previous results about BCP$_2$ are as follows. 
The BCP$_2$ on grid graphs of more than two rows was shown to be NP-hard \cite{bec98} while for grid graphs of two rows (also known as ``ladders''), the problem can be solved in polynomial time \cite{bec01}. 
Approximation algorithms for BCP$_q$ on grid graphs were also presented by \cite{bec98} but the general
approximation ratios were not given, except for the case $q = 2$, for which a
3/2-approximation can be guaranteed. 
Besides the ladders, it is known that the BCP$_q$ problem is
polynomially solvable for trees \cite{per81} and unweighted $q$-connected graphs \cite{lov77}.
\cite{chl96}
showed that BCP$_2$ on general graphs is NP-hard in the strong sense and cannot be
approximated with an absolute error guarantee of
$|V|^{1-\varepsilon}$ for any $\varepsilon>0$ unless NP=P. 
A $4/3$-approximation algorithm was also given in that paper, which is currently the best approximation ratio of the problem, even on grid graphs. 
For BCP$_3$ and BCP$_4$, on 3- and 4-connected graphs respectively, there are 2-approximation algorithms proposed by \cite{cha07}.

The rest of the paper is organized as follows: In Section 2, we give some notations and show that the min-NSC is a minimum non-separating path in a grid graph. In Section 3, we show the algorithm for the minimum non-separating path.
The 5/4-approximation algorithm for GBCP$_2$ is given in Section 4.
The exact algorithm and the approximation scheme of GBCP$_2$ are in Sections 5 and 6, respectively.
Finally, some concluding remarks are given in Section 7.

\section{Preliminaries}

Let $G=(V,E)$ be a graph, $S\subset V$ and $H$ an induced subgraph of $G$. 
The subgraph induced by $S$ is denoted by $G[S]$.
By $G-S$ we denote $G[V-S]$, and similarly $G-H=G[V-V(H)]$, in which $V(H)$ is the node set of $H$.
Let $w: V\rightarrow Z^{+}$ be a node weight function. 
By $w(S)$, we denote the total weight of $S$, i.e., $w(S)=\sum_{v\in S}w(v)$. 
For convenience $w(H)=w(V(H))$. 
Let $[i,j]$ denote the interval of integers $\{i,i+1,\ldots, j\}$ for $i\leq j$.
An $m\times n$ grid graph $M$ will be thought of as an $m\times n$ matrix such that $V(M)=\{M_{ij}| i\in [1,m], j\in [1,n]\}$ and there exists an edge between two consecutive nodes in the same row or the same column.  
The set of nodes in the first or the last row and the first or the last column is called as the {\em boundary} of the grid graph. 
The four nodes $M_{1,1}$, $M_{1,n}$, $M_{m,1}$ and $M_{mn}$ are called as corner nodes.
W.l.o.g. we assume $n\geq m$. 
Let $N=mn=|V(M)|$ and $w_{ij}$ the weight of node $M_{ij}$.

\begin{figure}[t]
\begin{center}
\epsfbox{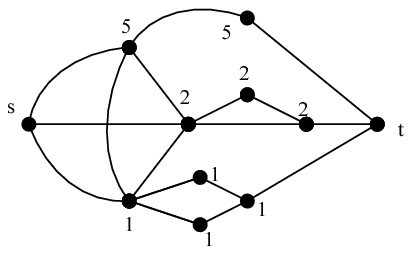}
\caption{The minimum NSC and the minimum non-separating (induced) $st$-path. The nodes are labeled by their weights. The minimum $st$-NSC consists of the four nodes of weight 1 (in addition to $s$ and $t$); the minimum non-separating $st$-path passes through the three nodes of weight 2; and the minimum non-separating induced $st$-path passes through the two nodes of weight 5.}
\label{borderpath}
\end{center}
\end{figure}

Let $\mathcal{C}_1$, $\mathcal{C}_2$ and $\mathcal{C}_3$ be the sets of all non-separating induced $st$-paths, non-separating $st$-paths and $st$-NSCs, respectively. By definition, $\mathcal{C}_1\subset \mathcal{C}_2\subset \mathcal{C}_3$.
For general graphs, they are different and there may be even no any non-separating $st$-path. 
Figure~\ref{borderpath} illustrates a case that the minimum NSC and the non-separating (induced) $st$-path are  different.
In the remaining of this section we show that $\mathcal{C}_1=\mathcal{C}_2=\mathcal{C}_3$ on a grid graph
except that $\{s,t\}$ is a \emph{2-cut}. 

\begin{figure}[t]
\begin{center}
\epsfbox{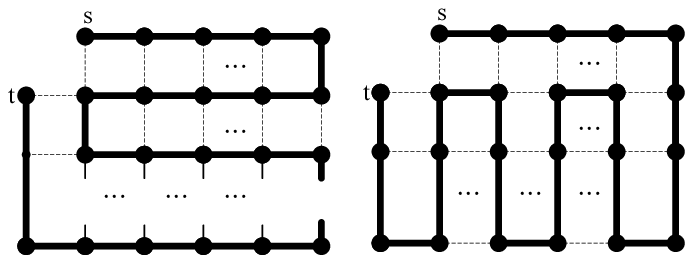}
\caption{The Hamiltonian $st$-paths in $G-x$ for even number of rows and even number of columns are shown in the left  and the right, respectively}
\label{hamil}
\end{center}
\end{figure}

A node subset is a {\em cut} if the graph becomes disconnected after its removal.
A 2-cut is a cut consisting of two nodes. 
For a grid graph of at least three rows, the only 2-cuts are the pairs of the two neighbors of corner nodes.
Suppose that $\{s,t\}$ is a 2-cut of an $m\times n$ grid graph $G$, in which $n\geq m\geq 3$.
Let $x$ be the corner node adjacent to both $s$ and $t$.
Apparently the minimum $st$-NSC is either the path $(s,x,t)$ or $G-x$, depending on which weight is smaller.
The minimum non-separating path is similar but a little tricky.
It is not hard to observe that the minimum non-separating $st$-path is either $(s,x,t)$ or a Hamiltonian $st$-path of $G-x$ if the Hamiltonian path exists. 
If both $m$ and $n$ are odd integers, we can show that there does not exist a Hamiltonian $st$-path in $G-x$ as follows. First we color $s$ white, and then all other nodes are colored according to the rule: ``the neighbors of a white node should be color black and vice versa''. It can be easily checked that $t$ is colored white and that the numbers of white and black nodes are the same. Since a path with both endpoints colored white cannot have equal number of white and black nodes, no Hamiltonian $st$-path exists on $G-x$.
On the contrary, if $m$ or $n$ is even, we can show that a Hamiltonian $st$-path always exists. 
W.l.o.g we assume that $x$ is the corner node $M_{1,1}$.
The Hamiltonian paths for even number of rows, and columns respectively, are illustrated in Fig.~\ref{hamil}. 

Anyway, if $\{s,t\}$ is a 2-cut, both the min-NSC and the minimum non-separating path can be easily computed, and 
we shall assume it is not the case in the remaining of the paper.
For an NSC $B$ or a connected partition $(V(B),V-V(B))$, a node $v$ is movable if 
it is still a connected bipartition after moving $v$ to the other part. 
A trivial observation is that $v\in V(B)$ is movable iff $v$ is not an articulation node in $B$ and $v$ has a neighbor in $V-V(B)$, assuming $V-V(B)$ is not empty.
For a connected bipartition of a biconnected graph, there are at least two movable nodes in each part unless the part contains less than two nodes \cite{chl96}. The following result comes from the minimality of the NSC.
\begin{lemma}\label{stmove}
If $B$ is a minimum $st$-NSC of a biconnected graph, both $s$ and $t$ are movable and there is no other movable node in $B$.
\end{lemma}

\begin{theorem}\label{inducepath}
If $\{s,t\}$ is not a 2-cut, a minimum $st$-NSC $B$ on a grid graph $M$ of at least three rows is a non-separating induced $st$-path.
\end{theorem}
\begin{proof}
By definition $M-B$ is connected, and it is sufficient to show that $B$ is an induced $st$-path.
A block is either a maximally biconnected subgraph or a bridge (an edge whose removal disconnects the graph).
Let $\mathcal{K}$ and $A$ be the sets of the blocks and the cut vertices of $B$, respectively. 
The \emph{block-cutpoint tree} $T$ of $B$ is defined as follows \cite[Chap. 4]{west}. 
$V(T)=\mathcal{K}\cup A$, and for any $a\in A$ and $K\in \mathcal{K}$, 
$(a,K)\in E(T)$ iff, in the original graph $B$, $a$ is a vertex in block $K$.
By definition $T$ is a tree and each leaf of $T$ corresponds to a block of $B$.

Since $M$ is biconnected, for any leaf $K$ of $T$, there is a node in the block $K$ of $B$ which is adjacent to $M-B$ and therefore movable.
By Lemma \ref{stmove}, $s$ and $t$ are the only movable vertices in $B$. 
So $T$ has at most two leaves, and is therefore a path.
We shall show that each block is an edge, and the proof is completed. 
Suppose by contradiction that $K$ is a block of $B$ and $|V(K)|>2$.
Since $M$ is a grid graph of at least three rows, there are at least three nodes in $K$ adjacent to $M-K$ unless $K$ contains all nodes but a corner (in this case $K=B$). If $K$ contained all nodes but a corner node, then $s$ and $t$ would be the two neighbors of the corner node and therefore a 2-cut, violating the assumption. 
A node in $K$ and adjacent to $M-K$ is either
\begin{itemize}
\item a cut of $B$ and therefore adjacent to a component of $B-K$; or
\item adjacent to $M-B$ and therefore a movable node of $B$ because no node of a block of more than two nodes is a cut of the block.
\end{itemize}
Since $M$ is biconnected, there is a movable node of $B$ in each component of $B-K$. 
Therefore there are at least three movable nodes in $B$, and by Lemma~\ref{stmove}, $B$ is not a minimum $st$-NSC.
\qed \end{proof}

\begin{corollary}
For a grid graph of at least three rows, the minimum NSC, the minimum non-separating path and the minimum non-separating induced path are all equivalent except that $s$ and $t$ are the two neighbors of a corner node.
\end{corollary}

\section{Minimum non-separating path}

In this section we show how to find a minimum non-separating path efficiently.
A boundary path is a path along the boundary, i.e., all the nodes are on the boundary.
A interior path is a path whose internal nodes are not on the boundary.
In the following we shall denote by $\mathcal{B}$ the boundary of $M$.
For a path $P$, a \emph{boundary subpath} of $P$ is a \emph{maximal} subpath of $P$ with all nodes on the boundary, 
i.e. a boundary subpath of $P$ is not a subpath of another boundary subpath of $P$. 
A subpath may contain only a single node. 

\begin{lemma}\label{subpath}
A non-separating induced $st$-path has at most one boundary subpath.
\end{lemma}
\begin{proof}
If $P$ is an induced path and has more than one boundary subpaths, there are at least two boundary segments divided by $P$, and each of these segments is in one component of $M-P$, which implies $P$ is not non-separating.
\qed \end{proof}

\begin{lemma}\label{onetouch}
If $P$ is a minimum $st$-path with at most one boundary subpath, then $P$ is a non-separating induced path.
\end{lemma}
\begin{proof}
Let $P=(s=v_1,v_2,\ldots, v_l=t)$. 
First we show that $P$ is an induced path by contradiction. 
If not, there exist $v_i$ and $v_j$ such that $(v_i,v_j)\in E$ and $i<j-2$.
Since removing a subpath from a path will not increase the number of boundary subpaths, we can replace the subpath from 
$v_i$ to $v_j$ with the edge $(v_i,v_j)$ to obtain a path of less weight and with at most one boundary subpath, a contradiction to the minimality of $P$. 

For any induced path, the subgraph induced by any of its node subset is not a cycle.
Therefore $P$ cannot include the whole boundary. 
Since $P$ has at most one boundary subpath, the remaining boundary nodes $\mathcal{B}-V(P)$ are connected and not empty.
If $P$ was not non-separating, there would be an interior node separated from the remaining boundary by $P$.
Since there is no induced cycle in $V(P)$ and $P$ only has at most one boundary subpath, 
it only happens at the case that the remaining boundary contains only a corner node, which implies that $P$ is a boundary path including the whole boundary but the corner node, i.e., $s$ and $t$ are the two neighbors of the corner node.  
But this contradicts the assumption that $\{s,t\}$ is not a 2-cut.
\qed \end{proof}

By the above results, a minimum NSC is a minimum non-separating path, and is also a minimum $st$-path with at most one boundary subpath. This property is helpful for designing our algorithm.
The next corollary is immediate.
\begin{corollary}\label{2boundary}
If both $s$ and $t$ are on the boundary, a minimum non-separating $st$-path is a boundary path and can be found in linear time to $|\mathcal{B}|$.
\end{corollary}

For two nodes $u$ and $v$, let $d_B(u,v)$, and $d_I(u,v)$, denote the minimum weight of any boundary path, and interior path respectively, between $u$ and $v$. For $d_B(u,v)$, both $u$ and $v$ must be on the boundary.
Let $d(u,v)$ denote the minimum weight of any non-separating induced $uv$-path.
We can have the next lemma.
\begin{lemma}\label{1boundary}
If $s$ is on the boundary and $t$ is not, 
$d(s,t)=\min_{i\in \mathcal{B}}\{d_B(s,i)+d_I(i,t)-w(i)\}$.
\end{lemma}
\begin{proof}
By Lemma~\ref{onetouch}, it is sufficient to compute the minimum weight of any $st$-path with at most one boundary subpath. Since $s$ is on the boundary, the optimal path must be a concatenation of a boundary subpath and an interior subpath, including the degenerating case that the boundary subpath is only one node, i.e., $s$ in this case.
\qed \end{proof}

For each $u$ of the four corner nodes, the value $d_I(v,u)=\infty$ for any $v$. 
It is trivial that the total time complexity to compute $d_B(s,i)$ for every $i\in \mathcal{B}$ is linear to $|\mathcal{B}|$. 
To compute $d(s,t)$, it is sufficient to find $d_I(t,i)$ for every $i\in \mathcal{B}$, and then the minimum can be found in $O(|\mathcal{B}|)$ time.
Since any boundary node $i$ other than a corner node has only one interior neighbor, $d_I(t,i)$ is the minimum weight of any path between $t$ and the interior neighbor of $i$ on $M-\mathcal{B}$. Since a minimum weight path can be found by an algorithm similar to Dijkstra's algorithm and the number of edges in a grid graph is linear to the number of nodes, the time complexity is dominated by calling to Dijkstra algorithm, which takes time $O(N\log N)$ \cite{dijk1}.

\begin{corollary}
If $s$ is on the boundary and $t$ is not, then $d(s,t)$ can be found in time $O(N\log N)$.
\end{corollary} 
The proof of the next result is similar to Lemma~\ref{1boundary} and is omitted. 
\begin{lemma}
If neither $s$ nor $t$ is on the boundary, then $d(s,t)$ is the minimum between $d_I(s,t)$ and 
\begin{eqnarray}
\min_{i,j\in \mathcal{B}}\{d_I(s,i)+d_B(i,j)+d_I(j,t)-w(\{i,j\})\} \label{path1}
\end{eqnarray}
\end{lemma}

Note that if the optimal path contains a boundary node, it must contain at least two boundary nodes since a boundary node has at most one interior neighbor.
The time for computing $d_I(s,t)$ is $O(N\log N)$ by Dijkstra algorithm.
To compute (\ref{path1}), a naive algorithm of checking all possible $i$ and $j$ takes quadric time. We shall give an algorithm with time complexity $O(|\mathcal{B}|)$ in the following.

Starting at an arbitrary boundary node, we number the boundary nodes clockwise from $1$ to $|\mathcal{B}|$. Let $w(i)$ be the weight of $i\in \mathcal{B}$ and $W_B$ denote the total weight of boundary nodes. 
Our algorithm has two rounds. In the first round, we find for every $i$ the best $j$ such that the minimum boundary path from $i$ to $j$ is clockwise. The other case that the path is counterclockwise is checked in the second round.
Since the two rounds are similar, we shall only show the first round.
For convenience, we double the whole sequence, i.e., 
the $(|\mathcal{B}|+i)$-th node is the same as the $i$-th node for $1\leq i\leq |\mathcal{B}|$.
The minimum boundary $ij$-path is clockwise if 
\[ \sum_{i<k<j}w(k)\leq \frac{1}{2}(W_B-w(i)-w(j)). \]
Therefore we define ${\rm right}(i)$ as the maximum index $j$ in $[i+1,i+|\mathcal{B}|-1]$ such that 
the path is clockwise.

Since ${\rm right}(i)$ is a increasing function, it is not hard to show that computing ${\rm right}(i)$ for every $i$ can be done in total time $O(|\mathcal{B}|)$. For every $i$ from 1 to $|\mathcal{B}|$, we find $j^*\in (i,{\rm right}(i)]$ minimizing $d_I(s,i)+\sum_{i<k<j^*}w(k)+d_I(t,j^*)$.
For a fixed $i$, equivalently we only need to find $j^*$ minimizing  
$\sigma(j^*)=\sum_{1\leq k<j^*}w(k)+d_I(t,j^*)$. 
By this way we reduce our problem to a \emph{range minimum query} (RMQ) problem. For a one dimensional array, there is an algorithm which reports the minimum in any index range in constant time after a linear-time preprocessing \cite{rmq}. 
For our problem, the array $\sigma$ has $2|\mathcal{B}|$ elements and we need to perform $|\mathcal{B}|$ queries. Therefore the time complexity for finding indices $i$ and $j$ minimizing (\ref{path1}) is $O(|\mathcal{B}|)$.
Clearly $\sigma(j)$ for every $j$ can be found in $O(N\log N)$ time.
Combining this result with Corollaries~\ref{2boundary} and \ref{1boundary}, we summarize this section in the next theorem. 

\begin{theorem}\label{nsptime}
A minimum non-separating path on an $N$-nodes grid graph can be found in $O(N\log N)$ time.
\end{theorem}

\section{A 5/4-approximation algorithm for GBCP$_2$}

In this section we show a 5/4-approximation algorithm with time complexity $O(N\log N)$ for the GBCP$_2$ problem by using minimum non-separating paths.
For the GBCP$_2$ problem, the optimal solution is trivial if there exists a node of weight at least $W/2$. Therefore we shall exclude this case in the following. 
We shall first introduce an algorithm for finding a bipartition of any biconnected graph but not necessarily of a grid graph. This algorithm is based on $st$-numbering and finds a 4/3-approximation of BCP$_2$. Then we show how to improve the ratio to 5/4 for grid graphs by using minimum non-separating paths.

\subsection{An algorithm based on $st$-numbering}

For a biconnected undirected graph $G=(V,E)$ and $s,t\in V$, an $st$-numbering is a 1-to-1 labeling $\lambda: V\rightarrow [1,|V|]$ satisfying $\lambda(s)=1$, $\lambda(t)=|V|$; 
and, for each node $v\in V-\{s,t\}$, $v$ has a neighbor with label smaller than $\lambda(v)$ and also a neighbor with label larger than $\lambda(v)$.
For a biconnected graph, an $st$-numbering always exists and can be found in linear time \cite{stnum}. The original algorithm for $st$-numbering requires that $s$ and $t$ must be adjacent. But, if $(s,t)\notin E$, we can simply add edge $(s,t)$ to obtain an $st$-numbering.

\begin{algorithm*}[ht]
{\bf Algorithm} {\sc STN}\\
{\bf Input: }A biconnected graph $G=(V,E)$.\\
{\bf Output: }A connected bipartition $G$.
\begin{algorithmic}[1]
\State find two nodes $s$ and $t$ of the largest and the second largest weights, respectively;
\State compute an $st$-numbering $\lambda$; 
\State let $V=\{v_i|1\leq i\leq n\}$ such that $\lambda(v_i)=i$, $\forall i$;
\State find $k$ such that $w(V_k)\leq W/2$ and $w(V_{k+1})>W/2$, in which $V_k=\{v_i| 1\leq i\leq k\}$;
\If{$w(V_k)\geq W-w(V_{k+1})$}
\State $k^*\leftarrow k$;
\Else 
\State $k^*\leftarrow k+1$;
\EndIf
\State output $(V_{k^*},V-V_{k^*})$. 
\end{algorithmic}
\end{algorithm*} 

\begin{claim}
For any $1\leq k<n$, the bipartition $(V_k,V-V_k)$ is a connected bipartition.
\end{claim}
\begin{proof}
For any node $v\in V_k$, since each node has a neighbor with smaller $st$-number, there exists a path from $v$ to $s$ in $G[V_k]$. Consequently $G[V_k]$ is connected. It can be shown similarly that $G[V-V_k]$ is also connected.
\qed\end{proof}

\begin{lemma}\label{stn}
The algorithm {\sc STN} takes linear time. If $V_{k^*}$ is the solution produced by the algorithm, then 
$\min\{w(V_{k^*}),W-w(V_{k^*})\}\geq (W-w_3)/2$, where $w_3$ is the third largest node weight in $G$. 
\end{lemma}
\begin{proof}
Due to \cite{stnum}, the $st$-numbering can be computed in linear time.
It is trivial that all the other steps can also be done in linear time.

When $k^*=k$, i.e., $w(V_k)\geq W-w(V_{k+1})$, we have 
$\min\{w(V_{k^*}),W-w(V_{k^*})\}=w(V_{k})\geq W-w(V_{k+1})$.
Similarly, when $k^*=k+1$, i.e., $w(V_k)< W-w(V_{k+1})$, we have 
$\min\{w(V_{k^*}),W-w(V_{k^*})\}=W-w(V_{k+1})> w(V_k)$.
That is, in either case, we have $\min\{w(V_{k^*}),W-w(V_{k^*})\}\geq 
\max\{ w(V_{k}),W-w(V_{k+1})\}\geq (w(V_k)+(W-w(V_{k+1})))/2=(W-w(v_{k+1}))/2$. 
Since no node has weight larger than $W/2$, we have that $v_{k+1}$ is neither $s$ nor $t$.
Therefore $w(v_{k+1})\leq w_3$.
\qed\end{proof}

\subsection{A 5/4-approximation algorithm}
By Lemma~\ref{stn} and the analysis in \cite{chl96}, it can be shown that the algorithm {\sc STN} is a linear-time 4/3-approximation algorithm for the BCP$_2$ problem on any biconnected graph. Of course it works for grid graphs since a grid graph is biconnected. The remaining paragraphs of this section aims at improving the approximation ratio to 5/4.
In the remaining we assume $G$ is an $m\times n$ grid graph and $N=m\times n$, in which $n\geq m\geq 3$. 
Let $H=\{h_i|w(h_i)>W/5\}$ be the set of \emph{heavy} nodes. Clearly $|H|\leq 4$. 
We shall show how to find a 5/4-approximation solution for each possible value of $|H|$. The minimum non-separating path will play an important role in the case of $|H|=3$.

\begin{claim}
When $|H|\leq 2$, the algorithm {\sc STN} finds a $5/4$-approximation.
\end{claim}
\begin{proof}
In this case the third largest node weight $w_3$ is at most $W/5$.
By Lemma~\ref{stn}, the returned bipartition satisfies $\min\{w(V_{k^*}),W-w(V_{k^*})\}\geq (W-W/5)/2=(2/5)W$. 
Comparing with the trivial upper bound $W/2$, the ratio is 5/4. 
\qed \end{proof}

\begin{claim}
When $|H|=4$, the algorithm {\sc STN} finds a $5/4$-approximation.
\end{claim}
\begin{proof}
Since the nodes are arranged in a linear order by their $st$-numbers,
there always exists a bipartition, said $\mathcal{P}$, whose both parts contain exactly two heavy nodes and are of weight larger than $(2/5)W$.
By the optimality of $k^*$, the output is no worse than $\mathcal{P}$. 
\qed \end{proof}

Finally we consider the case that $|H|=3$. 
\begin{definition}
Let $G$ be a graph and $U$ a subset of nodes of $G$. The contracted graph $G/U$ is the graph obtained by combining all the nodes in $U$ by a new node $u$ and, for any $v\notin U$, the edge $(u,v)$ exists iff $v$ has a neighbor in $U$. For convenience, $G/S=G/V(S)$ for a subgraph $S$.
\end{definition}
\begin{lemma}\label{contract}
If $G$ is biconnected and $(U,V(G)-U)$ is a connected bipartition, then $G/U$ is biconnected.
\end{lemma}
\begin{proof}
Since $(U,V(G)-U)$ is a connected bipartition, the new node $u$ is not a cut node in $G/ U$. Since $G$ is originally biconnected, no node in $V(G)-U$ will be a cut node in the contracted graph.
\end{proof}

Our 5/4-approximation algorithm for the case of three heavy nodes is as follows.

\begin{algorithm*}[ht]
{\bf Algorithm} {\sc Three\_Heavy}
\begin{algorithmic}[1]
\State let $H=\{h_1,h_2,h_3\}$ be the set of nodes of weight larger than $W/5$;
\State find a minimum $h_ih_j$-NSC $B_{ij}$ for each $1\leq i<j\leq 3$; 
\State let $B$ be the NSC of smallest weight among those found in Step 2; 
\If{$w(B)<W/2$}
\State call algorithm {\sc STN} on the contracted graph $G/ B$;
\Else 
\State output $(V(B),V-V(B))$.
\EndIf
\end{algorithmic}
\end{algorithm*} 

\begin{claim}\label{3heavy}
In the case of $|H|=3$, the algorithm {\sc Three\_Heavy} finds a 5/4-approximation in $O(N\log N)$ time.
\end{claim}
\begin{proof} 
By Lemma~\ref{contract}, the contracted graph $G/ B$ is biconnected and has exactly two nodes of weight larger than $W/5$.
If $w(B)<W/2$, the result is the same as the case of $|H|=2$.
For otherwise we claim that $(V(B),V-V(B))$ is an optimal connected bipartition.
Let $(U,V-U)$ be an optimal connected bipartition.
Since there are three heavy nodes, at least two heavy nodes must be in the same part.
W.l.o.g. we assume that heavy nodes $h_1$ and $h_2$ are in $U$.
By the definitions of NSC and $B$, we have that $w(U)\geq w(B_{12})\geq w(B)\geq W/2$, and this implies $(V(B),V-V(B))$ is an optimal connected bipartition.

By the result of previous section, $B$ must be an induced path or the whole grid graph lacking a corner node.
And in either case the contracted graph $G/ B$ can be easily constructed in linear time. 
Since the algorithm {\sc STN} takes also linear time, the total time complexity is dominated by the step of finding the minimum NSC's, which is $O(N\log N)$ according to Theorem~\ref{nsptime}.
 
\qed \end{proof}

We conclude this section as follows.
\begin{theorem}
The GBCP$_2$ can be approximated with ratio 5/4 in $O(N\log N)$ time.
\end{theorem}

\section{An exact algorithm for GBCP$_2$} 

In this section we develop an exact algorithm for GBCP$_2$.
We shall assume that the grid graph has at least three rows.
For $1\leq j\leq n$ and a bipartition $(V_0,V_1)$, 
we use a vector $\mathbf{z}_j\in \{0,1\}^m$ to represent a bipartition of the $j$-th column such that $\mathbf{z}_j^i=0$ if $M_{ij}\in V_0$ and is one otherwise, in which $\mathbf{z}_j^i$ denotes the $i$-th component of $\mathbf{z}_j$ for $1\leq i\leq m$. 

We shall represent by configuration $(\mathbf{z},\theta,\tau)$ a possible bipartition $(V_0,V_1)$ of the first $j$ columns such that the partition of the $j$-th column is $\mathbf{z}$, the weight of $V_1$ is $\theta$, and $\tau$ represents how the nodes of column $j$ are connected in the first $j$ columns. That is, for the first $j$ columns, if we delete the nodes of $V_1$, $\tau$ represents the connected components of the nodes of $V_0$; and similarly the components of $V_1$ if we delete the nodes of $V_0$. We say $\tau$ is a {\em connection topology}. 
We shall develop a dynamic programming algorithm computing all possible configurations for $j$ from 1 to $n$.
We first discuss the connection topology.

\subsection{Connection topology}

To represent a connection topology, it is sufficient to use a data structure for disjoint sets. Each set contains the row indices of the nodes in the same connected component. Precisely speaking, at column $j$, we use an array $\tau$ such that $\tau[i]$ stores the representer of the set $M_{ij}$ belongs to, in which the representer of a set is the smallest row index of its members. For example, if, at the first $j$ columns, the nodes $M_{2,j}$, $M_{3,j}$ and $M_{9,j}$ are in the same component, we record the set $\{2,3,9\}$ by 
$\tau[2]=\tau[3]=\tau[9]=2$.
Since the nodes in a component are all in $V_0$ or $V_1$, we divide a connection topology into  
a 0's connection topology, or 0-topology for short, and a 1-topology. 
That is, a $q$-topology is a partition of $\{i|\mathbf{z}_j^i=q\}$ for column $j$. A trivial upper bound of the number of connection topologies is $m^m$, and 
we aim at finding a much sharper bound. We shall first consider the 0-topology. 

First of all, if two adjacent nodes are in $V_0$, they must be in the same subset. We transform a column vector into a set of disjoint segments of 0's (0-segments for short). Labeling the segments by consecutive integers from 1, we have an interval $[1,p]$. Clearly $p\leq \lceil m/2\rceil$. For the above example of $\{2,3,9\}$, 2 and 3 are the first segment and 9 is the second segment. We shall represent the bound by a function of $p$. In the remaining a 0-topology shall be thought of as a partition of $[1,p]$. 

Consider the points $\{(i,0)|1\leq i\leq p\}$ in the Euclidean plane. The number of 0-topologies is the number of ways of adding some lines on one half-plane to joint some of these points. If two lines are cross, the corresponding segments are in the same component. 
\begin{definition}
For two subsets $S_1$ and $S_2$ in a 0-topology $\mathcal{T}$, $S_1$ is \emph{covered} by $S_2$ if $\max(S_2)>\max(S_1)$ and $\min(S_2)<\min(S_1)$, in which $\max(S_1)$ and $\min(S_1)$ denote the maximum and the minimum elements in $S_1$, respectively. A subset is \emph{covered} if it is covered by some other subset and is \emph{uncovered} otherwise. 
\end{definition}
A crucial observation is as follows:

\begin{quote}
If $\mathcal{T}$ is a 0-topology of $[1,i]$ and $i\in S\in \mathcal{T}$,
then $S-\{i\}$ cannot be covered by any $S'\in \mathcal{T}$.
\end{quote}

\begin{figure}[t]
\begin{center}
\epsfbox{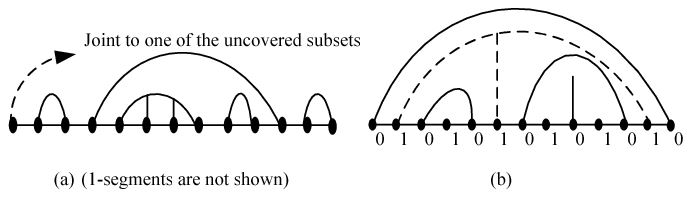}
\caption{Connection topologies (the columns are transposed).}
\label{topo}
\end{center}
\end{figure}

Let $\alpha(i,j)$ denote the number of 0-topologies of $i$ segments with $j$ uncovered subsets for $1\leq j\leq i\leq p$. In any 0-topology of $i$ segments, $i$ must be in one of the uncovered subsets of a 0-topology of $i-1$ segments. For a 0-topology of $i-1$ segments with $k$ uncovered subsets, joining $i$ to the $j$-th uncovered subsets will form a 0-topology of $i$ segments with $j$ uncovered subsets, for $j\leq k\leq i-1$. Besides, leaving $i$ itself as an uncovered subset along with a $j-1$ uncovered subsets also forms a 0-topology of $j$ uncovered subsets, 
seeing Fig.~\ref{topo}.(a). 
Therefore we can have the following recurrence relation.
\begin{eqnarray}
\alpha(i,j)=\sum_{k=j-1}^{i-1}\alpha(i-1,k) &{\rm for }\ 1\leq j<i \label{top-eq}
\end{eqnarray}
The boundary conditions are $\alpha(i,0)=0$ and $\alpha(i,i)=1$ for any $i$.

\begin{lemma}
For a fixed 0-topology with $j$ uncovered subsets, there are at most $2^j$ 1-topologies.
\end{lemma}
\begin{proof}
Clearly the 0-segments interleave the 1-segments, and vice versa.
Let $\mathcal{T}$ be a 0-topology. 
For $S\in\mathcal{T}$, we say that a 1-segment $i$ is \emph{minimally covered} by $S$ iff $i$ is covered by $S$ but not covered by any $S'\in\mathcal{T}$ covered by $S$. 
As shown in Fig.~\ref{topo}.(b), for any $S\in \mathcal{T}$, all the 1-segments minimally covered by $S$ is in the same component.
So, for the 1-segments covered by at least one subsets in $\mathcal{T}$, their components are fixed.
Therefore the number of 1-topologies only depends on the uncovered 1-segments. If there are $j$ uncovered subsets in $S$, there are at most $j+1$ uncovered 1-segments. 
Let $f(k)$ denote the maximum number of 1-topologies with $k$ uncovered 1-segments. Clearly $f(1)=1$.
For $k>1$, the $k$-th 1-segment either forms a component by itself or joins to the $(k-1)$-th 1-segment. 
Note that since they are uncovered, it cannot join to other segment but skip from the $(k-1)$-th one.
Consequently we have $f(k)=2f(k-1)$ for $k>1$, and it turns out to be $f(k)=2^{k-1}\leq 2^j$.
\qed \end{proof}

\begin{corollary}\label{n-top1}
For $p$ 0-segments, the number of connection topologies is upper bounded by $t(p)=\sum_{1\leq j\leq p}2^j\alpha(p,j)$.
\end{corollary}

\begin{lemma}\label{top-eq2}
For any $h\geq 0$ and positive $i$, $\sum_{k=0}^{h}{i-1+k\choose i-1}\leq {i+h\choose i}$.
\end{lemma}
\begin{proof}
By induction on $h$. First when $h=0$, $\sum_{k=0}^{h}{i-1+k\choose i-1}={i-1\choose i-1}={i+h\choose i}$.
Suppose that the inequality holds for $h-1$, i.e., $\sum_{k=0}^{h-1}{i-1+k\choose i-1}\leq {i+h-1\choose i}$. We have 
\begin{eqnarray*}
&&\sum_{k=0}^{h}{i-1+k\choose i-1}
=\sum_{k=0}^{h-1}{i-1+k\choose i-1}+{i-1+h\choose i-1}\\
&\leq& {i+h-1\choose i}+{i+h-1\choose i-1}
={i+h\choose i}
\end{eqnarray*}
\qed \end{proof}

\begin{lemma}\label{top-eq3}
$\alpha(i,j)\leq {2i-j\choose i}$ for $1\leq j\leq i$.
\end{lemma}
\begin{proof}
By induction on $i$.
When $i=1$, the only feasible value of $j$ is 1, and $\alpha(1,1)=1= {2i-j\choose i}$.
Suppose that it holds for $i-1$ and any $1\leq j\leq i-1$. By (\ref{top-eq}),
\begin{eqnarray*}
\alpha(i,j)&=&\sum_{k=j-1}^{i-1}\alpha(i-1,k) 
\leq \sum_{k=j-1}^{i-1}{2(i-1)-k\choose i-1}
=\sum_{k=0}^{i-j}{i-1+k\choose i-1}
\end{eqnarray*}
The last equality is obtained by substituting $k$ by $i-1-k$.
Then, by Lemma~\ref{top-eq2}, $\alpha(i,j)\leq {2i-j\choose i}$. 
\qed \end{proof}

\begin{lemma}\label{ntop}
$t(p)\leq 2^{2p}-{2p\choose p}$ for $p\geq 1$.
\end{lemma}
\begin{proof}
By Corollary~\ref{n-top1} and Lemma~\ref{top-eq3}, 
\begin{eqnarray*}
t(p)&=&\sum_{1\leq j\leq p}2^j\alpha(p,j)\leq \sum_{1\leq j\leq p}2^j{2p-j\choose p}
=\sum_{j=0}^{p-1}2^{p-j}{p+j\choose j}\equiv \hat{t}(p)
\end{eqnarray*}
The last equality is obtained by substituting $j$ with $p-j$.
We derive a closed form of $\hat{t}(p)$ by the method of generating function. 
First, $2^{p-j}{p+j\choose j}$ is the coefficient of $x^j$ in $2^{-j}(2+x)^{p+j}$, 
and is the coefficient of $x^p$ in $2^{-j}x^{p-j}(2+x)^{p+j}$.
Then $\hat{t}(p)$ is the the coefficient of $x^p$ in the following generating function:
\begin{eqnarray}
T(x)= \sum_{j=0}^{p-1}2^{-j}x^{p-j}(2+x)^{p+j} \label{eq1}
\end{eqnarray}
Since 
\begin{eqnarray}
\frac{2x}{2+x}T(x)&=&\sum_{j=0}^{p-1}2^{-j+1}x^{p-j+1}(2+x)^{p+j-1} \nonumber\\
&=&\sum_{j=-1}^{p-2}2^{-j}x^{p-j}(2+x)^{p+j} \label{eq2}
\end{eqnarray}
subtracting (\ref{eq2}) from (\ref{eq1}) and solving $T(x)$, we obtain 
\begin{eqnarray}
T(x) 
&=&\left(\frac{1}{2-x}\right)\left(2^{1-p}x(2+x)^{2p}-2x^{p+1}(2+x)^{p}\right) \nonumber\\
&=&\left(\sum_{i=0}^{\infty}\left(\frac{x}{2}\right)^i\right)\left(2^{-p}x(2+x)^{2p}-x^{p+1}(2+x)^{p}\right)
\nonumber
\end{eqnarray} 
Since $\hat{t}(p)$ is the coefficient of $x^{p}$ in $T(x)$, the second term is useless, and 
\begin{eqnarray*}
\hat{t}(p)&=&2^{-p}\sum_{i=0}^{p-1}\left(\frac{1}{2}\right)^i {2p\choose p-i-1} 2^{p+i+1} 
= 2\sum_{i=0}^{p-1}{2p\choose p-i-1} \\
&=& 2\sum_{i=0}^{p-1}{2p\choose i} 
= \sum_{i=0}^{2p}{2p\choose i}-{2p\choose p} 
=2^{2p}-{2p\choose p}  
\end{eqnarray*} 
\qed \end{proof}

Since $p$ is the number of 0-segments and bounded by $\lceil m/2\rceil$, we obtain the following result.  
\begin{theorem}
The number of connection topologies is $O(2^m)$.
\end{theorem}

\subsection{An exact algorithm for GBCP$_2$}

Let $L_j$ be the list of all configurations for column $j$.
Initially, for each binary $m$-vector $\mathbf{z}$, there is only one connection topology $\tau$ and only one possible total weight $\theta$ of $V_1$. So $L_1$ contains $2^m$ configurations.
For $j$ from 2 to $n$, our algorithm computes $L_j$ from $L_{j-1}$ in each iteration, as well as checks if a better feasible solution is obtained.
We say a component is ``closed'' at column $j-1$ if the component contains at least one node of column $j-1$ but none of column $j$.
For any $(\mathbf{z},\theta,\tau)\in L_{j-1}$ and any bipartition $\mathbf{z}'$ of column $j$, we check the following conditions:
\begin{itemize}
\item If $V_q$, $q\in \{0,1\}$, has exactly one component in $(\mathbf{z},\theta,\tau)$: 
\begin{itemize}
\item all nodes of column $\geq j$ assigned to $V_{1-q}$ is a feasible solution. We need only check if it is the currently best solution but not insert it into $L_j$.
\item any other $\mathbf{z}'$ closing the component is illegal and should be discarded.
\end{itemize} 
\item Otherwise any $\mathbf{z}'$ closing any component at column $j-1$ is illegal.
\item If it is not illegal, compute $\theta'$ and $\tau'$, and insert $(\mathbf{z}',\theta',\tau')$ into $L_j$.
\end{itemize}
Finally for each configuration of column $n$, we check if it is feasible and update the best one if necessary.
We next analyze the time complexity.

To store the configurations, we use a $2^m\times W$ table $T$ and each entry of the table is a list of connection topologies. For a configuration 
$(\mathbf{z},\theta,\tau)$, we store $\tau$ in the list of $T[\mathbf{z},\theta]$.
By using a balanced binary search tree (such as AVL tree) with array $\tau$ as the key, 
we can check the existence and insert a configuration in $O(m^2)$ time.
Note that we need to check and avoid the duplicates to ensure the number of configurations is bounded. 
The number of configurations is $O(2^m\times W\times 2^m)=O(W\times 4^m)$.
Since there are $n$ iterations and in each iteration it takes $O(m^2)$ time for every configuration and every $\mathbf{z}'$, 
the total time complexity is $O(m^2nW8^m)=O(mNW8^m)$. It is much better than $O(N2^N)$ of the brute force algorithm. 

\begin{theorem}\label{bcpexactime}
The GBCP$_2$ can be solved in $O(mNW8^m)$, in which $N$ is the number of nodes, $m$ is the number of rows, and $W$ is the total weight.
\end{theorem}
\section{Approximating GBCP$_2$}

Based on the exact algorithm for GBCP$_2$ and a scaling technique, we shall show an approximation algorithm for GBCP$_2$, which is a FPTAS for fixed number of rows. 

For some $\rho<1$, we scale down the weights by a factor 
$r=\rho W/(3N)$, i.e., we set $w'_{ij}=\lfloor w_{ij}/r \rfloor$ for each $i$ and $j$.
Then we run the exact algorithm on the modified weight $w'$ to obtain a bipartition $(V_0,V_1)$ as the approximation solution. Let $(V_0^*,V_1^*)$ be the optimal bipartition w.r.t. $w$.
The following result can be derived from Lemma~\ref{stn}.
\begin{lemma}\label{onethird}
If no node has weight larger than $W/2$, then $\min\{w(V_0^*),w(V_1^*)\}\geq W/3$.
\end{lemma}

\begin{lemma}\label{fratio}
$\frac{\min\{w(V_0^*),w(V_1^*) \}}{\min\{w(V_0),w(V_1) \}}\leq \frac{1}{(1-\rho)}$.
\end{lemma}  
\begin{proof}
W.l.o.g. let $w(V_0)\leq w(V_1)$. 
By definition, $w(V_0)\geq rw'(V_0)$. Since $(V_0,V_1)$ is the optimal bipartition w.r.t. $w'$, $\min_q w'(V_q)\geq \min_q w'(V^*_q)$ and therefore 
\[ w(V_0)\geq rw'(V_0)\geq r\min_q w'(V^*_q) \]

By the definition of $w'$, for $q=0$ or 1, $w(V^*_q)< rw'(V^*_q)+rN$. 
Hence,\\ $r\min_q w'(V^*_q)>\min_q w(V^*_q)-rN$, and we obtain
\[ w(V_0)>\min_q w(V^*_q)-rN \]

Then, since $r=\rho W/(3N)$ and $\min_q w(V^*_q)\geq W/3$ by Lemma~\ref{onethird}, the approximation ratio can be calculated by 
\begin{eqnarray*}
&&\min_q w(V^*_q)/w(V_0)\leq \frac{\min_q w(V^*_q)}{\min_q w(V^*_q)-rN}
\leq \frac{W/3}{W/3-\rho W/3}=\frac{1}{1-\rho},
\end{eqnarray*}
which completes the proof.
\qed \end{proof}

\begin{theorem}
For any $\varepsilon>0$, a $(1+\varepsilon)$-approximation of the GBCP$_2$ can be found in $O((1+\frac{1}{\varepsilon})mN^28^m)$ time, in which $N$ is the number of nodes and $m$ is the number of rows. For fixed number of rows, it is a FPTAS.
\end{theorem}
\begin{proof}
For any given $\varepsilon>0$, we set $\rho=\frac{\varepsilon}{1+\varepsilon}$ and $r=\rho W/(3N)$. By Lemma~\ref{fratio}, the approximation ratio is $1+\varepsilon$. By Theorem~\ref{bcpexactime}, the time complexity is $O(mNW'8^m)=O(mN(\frac{W}{\rho W/(3N)})8^m)=O((1+\frac{1}{\varepsilon})mN^28^m)$.

\qed \end{proof}

\section{Concluding remarks}

The technique in Section 4 may be used for other classes of graphs.
For any graph class on which the minimum non-separating path problem can be solved in polynomial time, the BCP$_2$ problem can be approximated with ratio 5/4 in the same time complexity. 

The approximation algorithm shown in Section 6 is an FPTAS only for fixed number of rows. 
The interesting open problems include how to design an FPTAS or PTAS for the GBCP$_2$ of non-fixed number of rows and how to evenly partition a (grid) graph into more than two connected subgraphs.

\section*{Acknowledgements}
This work was supported in part by 
NSC 97-2221-E-194-064-MY3 and NSC 98-2221-E-194-027-MY3 from the National Science Council, Taiwan.



\section*{Appendix: Erratum}

\begin{figure}[t]
\epsfbox{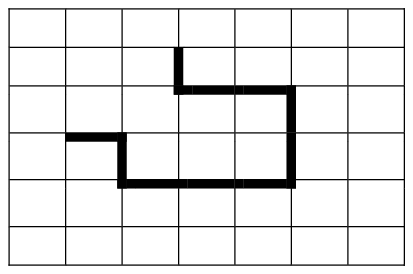}
\caption{A separating path without boundary subpath but having two subpaths in one face.}
\label{f1}
\end{figure}

The mistake appears in the proof of Lemma \ref{onetouch} which states that  
\begin{quote}
If $P$ is a minimum $st$-path with at most one boundary subpath, then $P$ is a non-separating induced path.
\end{quote}
However, Figure~\ref{f1} is a counter example. Due to this mistake, the algorithm for minimum non-separating paths is also wrong. 

The wrong algorithm for finding a minimum non-separating path is later used to find a 5/4-approximation for GBCP$_2$.
So far, we do not know how to fix the mistake for the non-separating path problem. Fortunately, we can fix the 5/4-approximation for GBCP$_2$ as follows. 
   
In this problem, we are given a vertex-weighted grid graph, i.e., a two-dimensional matrix, and the goal is to partition it into two connected subgraphs as balanced as possible. The cost function is defined by the smaller weight of the two subgraphs. By some previous results we only need to consider the case that both the numbers of rows and columns are at least three.

Let $W$ be the total weight of the graph and $H=\{h_i|w(h_i)>W/5\}$ be the set of \emph{heavy} nodes. Clearly $|H|\leq 4$. The 5/4-approximation algorithm is divided into cases for each possible value of $|H|$. The algorithm for minimum non-separating path is used in the case of $|H|=3$. We shall show how to find a 5/4-approximation solution for $|H|=3$.     
Let $H=\{h_1,h_2,h_3\}$ and $w_1\geq w_2\geq w_3$ be their weights, respectively. There are two cases.
\begin{itemize}
\item All the heavy nodes are on the boundary. The minimum non-separating for each pair of heavy nodes can be easily found, and therefore the previous algorithm works.
\item Otherwise, we claim that there are three internally disjoint paths $P_i$, $1\leq i\leq 3$, between $h_2$ and $h_3$ such that 
\begin{enumerate}
\item $P_1$ passes through $h_1$; and
\item $P_2$, $P_3$, and $Q$ are non-separating, where $Q$ is the subpath from $h_3$ to $h_1$ on $P_1$. 
\end{enumerate}
Since $G$ is 3-connected, there are three internally disjoint $h_2h_3$-paths. When both $h_2$ and $h_3$ are on the boundary, $P_2$ and $P_3$ are the two boundary paths, which are non-separating, and there must be an internal $h_2h_3$-path passing through $h_1$. When $h_2$ or $h_3$ is not on the boundary, we can easily find the three desired paths. 
\end{itemize}

If $w(Q)\leq 3W/5$, then $(Q,G-Q)$ is a 5/4-approximation solution since $Q$ contains two heavy nodes and has weight larger than $2W/5$. Otherwise, we have $w(Q)>3W/5$, and $q\equiv w(Q)-w_3>3W/5-w_3$. 
W.l.o.g. we assume that $w(P_2)\leq w(P_3)$, and we claim $D=(P_2,G-P_2)$ is a 5/4-approximation solution.

If $w(P_2)\leq W/2$, then we have done. Otherwise, the cost of the 2-partition $c(D)$ is  
\begin{eqnarray*}
W-w(P_2)&\geq& q+(W-q-w_2-w_3)/2\\
&=&W/2+q/2-(w_2+w_3)/2 \\
& >&W/2+3W/10-w_3/2-(w_2+w_3)/2 \\
&\geq& 4W/5-(w_2+w_3)/4-(w_2+w_3)/2\\
&=&4W/5-3(w_2+w_3)/4.
\end{eqnarray*}
We divide the proof into two cases: $w_2+w_3\leq W/2$ or $w_2+w_3> W/2$.
When $w_2+w_3\leq W/2$, we have that 
\[ c(D)\geq (4/5-3/8)W=17W/40>2W/5, \]
and the approximation ratio is less than 5/4. 
When $w_2+w_3> W/2$, the optimal cost is at most $W-(w_2+w_3)$, and the ratio is 
\[ \frac{W-(w_2+w_3)}{4W/5-3(w_2+w_3)/4}<20/17<5/4.\]
\end{document}